\documentstyle[12pt,psfig]{article}

\begin{document}
\begin{center}
{\large \bf Equilibrating Glassy Systems with Parallel Tempering}
\end{center}

{Walter Kob$^{*}$, Claudio Brangian$^{*}$,
Torsten St\"uhn$^{*}$, and Ryoichi Yamamoto$^{\dag}$}\\

\noindent
{$^*$ Institute of Physics, Johannes Gutenberg-University,
Staudinger Weg 7, \\
D--55099 Mainz, Germany\\
\noindent
$^{\dag}$ Department of Physics, Kyoto University, Kyoto 606-8502, Japan}
\begin{center}
16.03.2000
\end{center}

\begin{abstract}
We discuss the efficiency of the so-called parallel tempering method
to equilibrate glassy systems also at low temperatures. The main focus
is on two structural glass models, SiO$_2$ and a Lennard-Jones system,
but we also investigate a fully connected 10 state Potts-glass. By
calculating the mean squared displacement of a tagged particle
and the spin-autocorrelation function, we find that for these three
glass-formers the parallel tempering method is indeed able to generate,
at low temperatures, new independent configurations at a rate which is
$O(100)$ times faster than more traditional algorithms, such as molecular
dynamics and single spin flip Monte Carlo dynamics. In addition we find
that this speedup increases with decreasing temperature. The reliability
of the results is checked by calculating the distribution of the energy
at various temperatures and by showing that these can be mapped onto
each other by the reweighting technique.

\vspace{-0.3cm}

\end{abstract}

\section{Introduction}

Despite the progress made in recent years~\cite{vigo,mct}, our
understanding of the dynamics of deeply supercooled liquids and the
properties of glasses is still far from satisfactory~\cite{binder99}.
Although the so-called mode-coupling theory (MCT)~\cite{mct} seems to give
a very reliable description of this dynamics around the MCT-temperature
$T_c$, the relaxation dynamics significantly below $T_c$, i.e. close
to the experimental glass transition temperature $T_g$, still waits
for a deeper theoretical understanding. The problem is that in this
temperature range the mechanism leading to the relaxation of the system
seems to be governed by the so-called hopping processes, i.e. the
system relaxes in an activated way, and that these processes are not
readily describable by MCT in its ideal version. (Although the extended
version of MCT~\cite{mct2}, which to some extend takes into account these
hopping processes, is able to make precise predictions for the dynamics
at intermediate time scales, it looses its predictive power at long
times, i.e. the $\alpha-$relaxation regime.) Although experiments are
very useful to give information on {\it particle averaged} properties of
supercooled liquids and glasses, such as the viscosity or the intermediate
scattering function, they are less well adapted to study the system on
a very local scale, e.g.  on the level of a single particle. This is,
however, exactly the information which is needed if one wants to come to
an understanding of the relaxation mechanism in these systems if one wants
to go beyond the use of simple models, such as a particle moving around
in a frozen environment or similar simple pictures. One method which in
the past has been proven to be extremely useful to reveal the necessary
details of structure and dynamics of complex systems are computer
simulations~\cite{binder96}. Since such simulations allow to access at
any instant the full microscopic information, they are also ideally suited
to learn more about the relaxation mechanisms in supercooled systems and
glasses~\cite{kob99}.  One problem of such simulations is, however, that
the time and length scales accessible are rather limited (typically 100~ns
and box sizes of 100\AA$^3$) and this has prevented in the past to study,
e.g., the {\it equilibrium} dynamics of supercooled liquids close to
$T_g$, where the typical time scales are on the order of 10-1000 s, i.e.
15-17 decades longer than the typical time step of used in a simulation
which is on the order of 1/100 of the inverse of the Debye frequency,
i.e. 1-10~fs. One possibility to avoid this problem to some extend is
to use a sophisticated Monte Carlo algorithm that allows to equilibrate
the system even at those low temperatures at which standard molecular
dynamics or Monte Carlo algorithms fail to reach equilibration. In recent
years various authors have proposed different Monte Carlo algorithms
that should be suited to reach the equilibrium state even at low
temperatures~\cite{Berg,Lee,Lyubartsev,Marinari,Hukushima,Swendsen}
and have also been successfully applied to various
systems~\cite{Hansmann,Nakajima,Achenbach,Sugita,Yamamoto} (see, however,
Ref.~\cite{Bhattacharya} for discouraging results). Here we will discuss
the application of the so-called parallel tempering (PT) algorithm,
proposed by Hukushima {\it et al.}~\cite{Hukushima}, a method that
is similar in spirit to the replica exchange algorithm proposed earlier by
Swendsen and Wang~\cite{Swendsen}, to the case of structural glasses and
to the Potts-glass. Although for structural glasses with a small number
of particles ($N\leq 36$) the PT method has been successfully tested by
Coluzzi and Parisi~\cite{Coluzzi} we are here interested in using larger
systems ($N=336$ and $N=1000$). In the next sections we will discuss the
details of the PT algorithm, then introduce the models we investigated
and subsequently discuss the results.

\section{The parallel tempering algorithm}

The PT method proposed by Hukushima {\it et al.} can be summarized
as follows~\cite{Hukushima}: 1) We denote the Hamiltonian of
interest by $H=K(\bf{p})+E(\bf{q})$, where $K$ and $E$ are the kinetic and
potential energy, respectively, and ${\bf p}=(p_1,p_2,\ldots,p_N)$ and
${\bf q}=(q_1,q_2,\ldots,q_N)$ are the momenta and coordinates of the
particles, respectively. (If this is a spin-like Hamiltonian, we put
$K=0$.)  We now construct a new system consisting of $M$ noninteracting
subsystems, each composed of $N$ particles, with a set of
arbitrary particle configurations $\{{\bf q}_1,\cdots,{\bf q}_M\}$
and momenta $\{{\bf p}_1,\cdots,{\bf p}_M\}$.  The Hamiltonian of the
$i$-th subsystem is given by
\begin{equation}
H_i({\bf p}_i,{\bf q}_i)=K({\bf p}_i)+\Lambda_i E({\bf q}_i), 
\label{eq1}
\end{equation}
where $\Lambda_i\in\{\lambda_1,\cdots,\lambda_M\}$ is a parameter to scale
the potential. 2) A molecular dynamics simulation is done for the total
system, whose Hamiltonian is given by ${\cal H}=\sum_{i=1}^M H_i$, at a
constant temperature $T=\beta_0^{-1}$. In this way we obtain a canonical
distribution $P({\bf q}_1,\cdots,{\bf q}_M;\beta_0)= \prod_{i=1}^M P({\bf
q}_i;\Lambda_i\beta_0)\propto \exp[-\beta_0\sum_{i=1}^M \Lambda_i E({\bf
q}_i)]$ in configuration space. 3) After each time interval $\Delta
t_{PT}$, we attempt to exchange the potential scaling parameter of the
$m$-th and $n$-th subsystem, while $\{{\bf q}_1,\cdots,{\bf q}_M\}$ and
$\{{\bf p}_1,\cdots,{\bf p}_M\}$ are unchanged.  The acceptance of the
exchange is decided in such a way that it takes care of the condition
of detailed balance. Here we use the Metropolis scheme, and thus the
acceptance ratio is given by
\begin{equation}
w_{m,n}= \left\{
\begin{array}{ll}
1,&\qquad \Delta_{m,n}\le 0\\
\exp(-\Delta_{m,n}),&\qquad\Delta_{m,n}> 0,
\end{array}
\right.
\label{eq2}
\end{equation}
where $\Delta_{m,n}=\beta_0(\Lambda_n-\Lambda_m)(E({\bf q}_m)-E({\bf
q}_n))$. 4) If steps 2) and 3) are repeated for a sufficiently long time
this scheme leads to canonical distribution functions $P(E;\beta_i)$
at a set of inverse temperatures $\beta_i=\lambda_i\beta_0$. To make a
measurement at an inverse temperature $\beta_l$ one has to average over
all those subsystems $(i\in 1,\cdots,M)$ for which we have (temporarily)
$\beta_l=\lambda_i\beta_0$. 

Although the algorithm presented is correct for general choice of the
values of $m$ and $n$ in Eq.~(\ref{eq2}), it is advisable to exchange
only {\it neighboring} subsystems in order to allow for a reasonably high
acceptance rate. In addition one has also to choose the values of the
coupling constants $\{\lambda_1,\cdots,\lambda_M\}$ in such a way that
neighboring subsystem have an sufficiently large overlap in their
distribution of the energy.

We also note that the algorithm we just described shares many properties
with the one proposed by Swendsen and Wang in 1986~\cite{Swendsen}. These
authors already realized that it is very useful to connect dynamically
configurations at low temperatures to the ones at high temperatures and
also gave a specific method how this can be successfully done in the
case of a frustrated spin system.

\section{Models and Details of the Simulations}

In this paper we consider three different types of glassy systems: A
realistic model for SiO$_2$, one of the prototypical glass formers. A
binary mixture of Lennard-Jones particles, i.e. an example for a simple
glass former, and finally a ten state Potts-glass, an example for a
spin system with a discontinuous transition from the paramagnetic phase
to the spin glass phase.

SiO$_2$ is a prototype for a so-called {\it strong} glass
former~\cite{angell85}.  This means that the temperature dependence of
transport quantities like the viscosity or the diffusion constant show
an Arrhenius dependence. It is believed that this property is related
to the fact that the structure of amorphous silica is a open tetrahedral
network that is essentially independent of temperature.

The SiO$_2$ model we use has been proposed by van Beest {\it et al.} (BKS)
~\cite{beest91} on the basis of {\it ab initio} calculations.  In this
system two ions of type $\alpha$ and $\beta$ ($\alpha,\beta \in \{{\rm Si,
O}\})$ that are a distance $r$ apart interact via the following potential:
\begin{equation}
\phi(r)=
\frac{q_{\alpha} q_{\beta} e^2}{r} + 
A_{\alpha \beta} \exp\left(-B_{\alpha \beta}r\right) -
\frac{C_{\alpha \beta}}{r^6},
\label{eq3} 
\end{equation}
where the values of the parameters $q_{\alpha}$, $A_{\alpha
\beta}$, $B_{\alpha \beta}$, and $C_{\alpha \beta}$
can be found in Ref.~\cite{beest91}. Previous computer
simulations have shown that the BKS model gives a very
good description of the static and dynamic properties of real
silica~\cite{vollmayr96,koslowski97,horbach99a,horbach99b,jund99,horbach99c}.
However, so far these tests could be done only at relatively high
temperatures, i.e.  $T\geq 2750$K, since below this temperature the
relaxation time exceeds the time scale accessible to normal molecular
dynamics simulations (in this case 20~ns). This temperature has to be
compared with the experimental value of $T_g$, 1450K, i.e. so far it
has been possible to equilibrate the system only at temperatures two
times higher than $T_g$.

In the following we will present results in which we used the PT
algorithm to equilibrate the system at low temperatures. Between the
attempted exchanges between two subsystems we propagated the particles
in the isokinetic ensemble at constant volume. The system size was
336 ions and the time step was 1.6~fs. The results presented below are
for $M=32$ subsystems, the number of time steps for equilibration and
production was each $4\cdot 10^6$, and the value of $\Delta t_{PT}$
was 1000 time steps. More details on this simulation can be found in
Ref.~\cite{stuhn00}.

The second structural glass is a binary (80:20) mixture of
Lennard-Jones particles. If we denote the majority species by A and
the minority species by B the interaction between two particles
are given by $\phi_{\alpha\beta}(r)=4\epsilon_{\alpha\beta}
[(\sigma_{\alpha\beta}/r)^{12}-(\sigma_{\alpha\beta}/r)^6]$, where
$r$ is the distance between particles $i$ and $j$. The interaction
parameters are $\alpha,\beta\in \{{\rm A,B}\}$, $\epsilon_{\rm AA}=1$,
$\epsilon_{\rm AB}=1.5$, $\epsilon_{\rm BB}=0.5$, $\sigma_{\rm AA}=1$,
$\sigma_{\rm AB}=0.8$, and $\sigma_{\rm BB}=0.88$. In the following
we will measure length and energy in units of $\sigma_{\rm AA}$
and $\epsilon_{\rm AA}$, respectively, (setting $k_B=1$) and time
in units of $(m\sigma_{AA}^2/48\epsilon_{AA})^{1/2}$, where $m$ is
the mass of the particles. In the simulation we used a cubic box,
of length 9.4, with periodic boundary conditions, the total number
of particles was 1000, and the time step was 0.01732. The number of
subsystems was 16, the number of time steps for equilibration and
production was each $5\cdot 10^6$, and the value of $\Delta t_{PT}$
was 1000 time steps. More details on the simulation can be found in
Ref.~\cite{Yamamoto}. In the past many properties of this model have been
investigated~\cite{kob99,kob_lj,vollmayr96b,gleim98,sciortino99,donati99,kob00}
and it has been found that its relaxation dynamics becomes very slow at
around $T=0.45-0.43$, i.e. its starts to exceed the time scales accessible
to normal molecular dynamics simulations, which is on the order of $10^8$
time steps. In contrast to the Arrhenius-like increase of the relaxation
times, as it is found in SiO$_2$, this system shows at low temperatures
an increase which can be fitted well with a power-law, and thus this
is considered to be a {\it fragile} glass former~\cite{angell85}. The
structure of the system resembles the one of randomly closed packed hard
spheres and is thus very different from the open network in silica.
Last not least it has to be mentioned that the simplicity of the
interaction of this model allows to obtain results for this system which
have a significantly higher statistical accuracy than the ones for the
silica model discussed above, since in the latter one has to calculate
numerically expensive long range interactions. This is the reason why
in the results discussed below the data for the silica system is quite
a bit more noisy than the one for the Lennard-Jones system.

The Potts-glass we consider is an example of a spin glass which shows a
discontinuous transition from the paramagnetic phase to a spin glass phase
if the number of states is larger than 4~\cite{potts_glass}. It has been
suggested that such type of models show a qualitatively similar dynamics
as structural glasses~\cite{kirkpatrick89,parisi97}, and therefore it is
of interest to understand their properties in more detail. Here we use the
version in which each spin $\sigma_i$ can take one of $q=10$ different
states ($\sigma_i\in\{1,2,\ldots,10\}$) and each spin interacts with
every other one with an interaction $J_{ij}$.  Thus the Hamiltonian is:
\begin{equation}
H=-\frac{1}{2}\sum_{i}^N \sum_{j\neq i}^N J_{ij} \quad .
\left(q \delta_{\sigma_i,\sigma_j}-1\right)
\label{eq4}
\end{equation}
The interactions $J_{ij}$ are drawn from a gaussian
probability distribution with mean $(3-q)/(N-1)$ and variance
$(N-1)^{1/2}$\cite{potts_glass}. It has been show that in the
thermodynamic limit this model has a dynamical singularity at $T_d=1.14$,
slightly above the static singularity at $T_c=1.13$~\cite{santis95}.
We have considered system sizes $N$ between 32 and 2560~\cite{brangian00}
but here we will discuss only results for $N=320$.  In principle one has
of course to average all the results obtained over the quenched disorder,
i.e. the interactions $J_{ij}$. For the present case we have not done this
(i.e. we consider only one realization of the disorder) but we have
tested that the results presented below are independent of the choice
of $J_{ij}$.  The number of subsystems was $M=16$ and the exchange time
$\Delta t_{PT}$ was 10 Monte Carlo steps per spin. Below we will compare
the results of the relaxation dynamics with the PT algorithm with the one
of standard single spin flip Monte Carlo scheme. In the latter case we
used the Metropolis criterion to accept or reject a move. More details
on this investigation can be found in Ref.~\cite{brangian00}

\section{Results}

In this section we discuss the results, i.e. we compare the efficiency of
the PT algorithm with the more conventional methods (standard molecular
dynamics and Metropolis Monte Carlo) to propagate the system through
configuration space. First we present our findings for the SiO$_2$ system,
then for the Lennard-Jones system, and finally for the Potts-glass. In
all cases we first used the PT algorithm for a sufficiently long time
to allow all subsystems to equilibrate. This equilibration was tested by
comparing quantities like the energy or the specific heat obtained from
such runs with the results from similar runs done with a standard method
(molecular dynamics, single flip Monte Carlo). 

\begin{figure}[b]
\psfig{figure=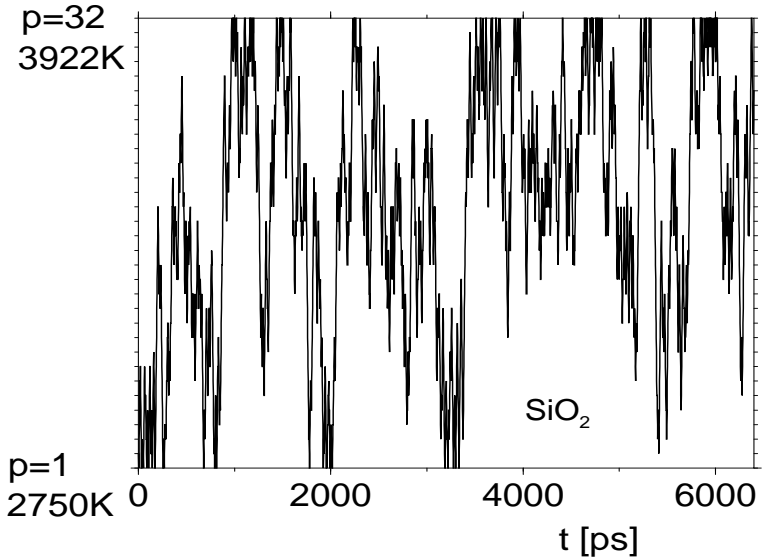,width=10cm,height=7.5cm}
\caption{Time dependence of the coupling constant of one subsystem for
the case of the SiO$_2$ system.}
\label{fig1}
\end{figure}

From the setup of the PT algorithm it follows that each subsystem makes
a random walk in temperature space. A rough estimate for the time needed
until a new low-temperature configuration is produced can be obtained
by looking at this random walk. For the case of SiO$_2$ a typical
example is shown in Fig.~\ref{fig1}, where we plot the temperature of
one subsystem, or rather the value of the coupling constant $\Lambda_i$
(see~(\ref{eq1})), as a function of time. The lowest temperature ($p=1$)
is 2750K and the highest one ($p=32$) is 3922K. From this figure we
recognize that it takes the system a bit more than 1000ps to generate a
new independent configuration, i.e. to go from $p=1$ to $p=M=32$ and back.
As we will see below this is significantly shorter than the time it takes
in a standard molecular dynamics simulation. We also mention that the
computational cost is of course also $M$ times (=number of subsystems) higher
than in a conventional scheme. However, it has to be remembered that we
also obtain $M$ independent configurations and not just one (this holds
for each temperature!). Therefore
if one needs many independent configurations, as it is usually the case
in order to obtain reliable averages, the PT algorithm will pay off. 

From this run we can calculate the distribution of the potential
energy at the various temperatures. These distributions are shown in
\begin{figure}[t]
\psfig{figure=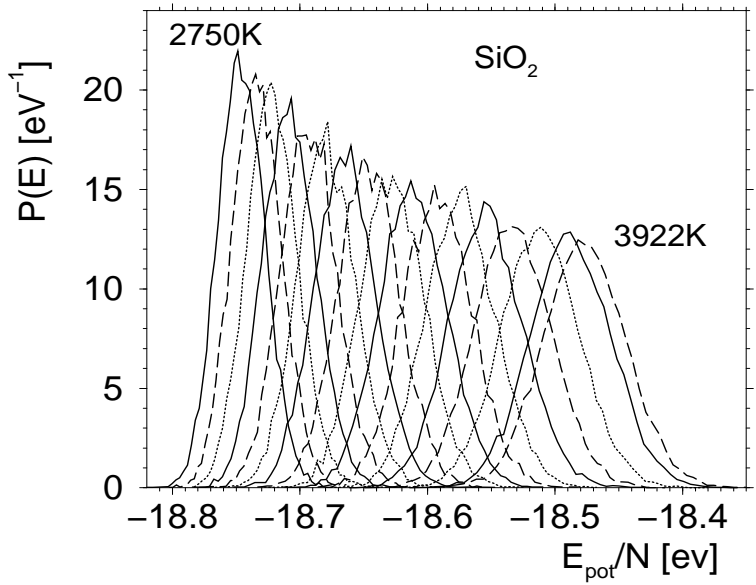,width=10cm,height=7.5cm}
\caption{Distribution of the energy for the different subsystems for the SiO$_2$
system.}
\label{fig2}
\end{figure}
Fig.~\ref{fig2}. (In order to avoid overcrowding of the figure we show
only every second temperature and the highest one.) We see that in order
to obtain a sufficiently high acceptance ratio it is necessary to have
a good overlap of neighboring distributions. A smaller overlap will lead
to a smaller acceptance probability and hence the random walk will take
longer. On the other hand a smaller value of $M$ will allow the random
walk to go faster from low temperatures to high temperatures. It is
presently not clear what the optimal choice is since this will depend on
the details of the system. More results on this can be found
in Ref.~\cite{stuhn00}.

At the beginning of this section we have mentioned that one possibility
to test whether or not the system has reached equilibrium is to compare
various observables with the ones obtained with a conventional simulation
method. This approach is, however, only possible for those temperatures
at which it is feasible to equilibrate the system also with one of the
latter methods. A different possibility is to use the data for the
energy distribution for the different subsystems and to test whether
\begin{figure}[t]  
\psfig{figure=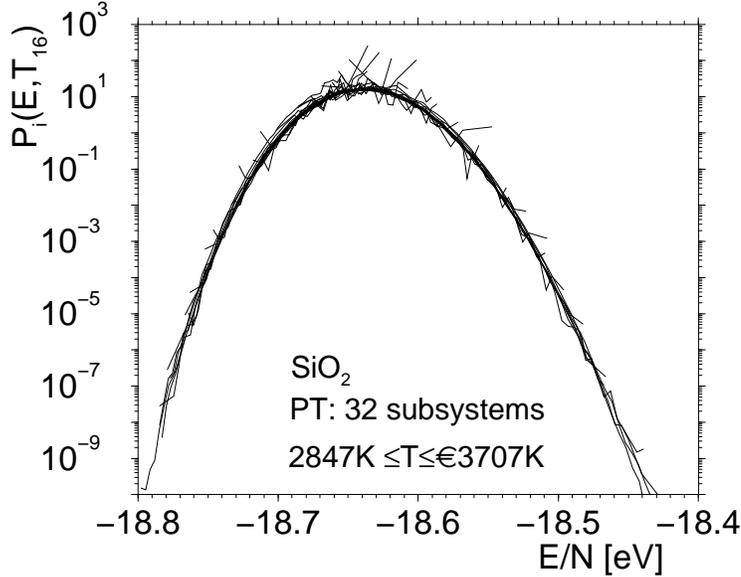,width=10cm,height=7.5cm}
\caption{Reweighted probability distributions for the energy of the SiO$_2$ system
for the temperature interval given in the figure. See main text for more details.}
\label{fig3}
\end{figure}
it is possible to reweight neighboring distributions to one common
temperature~\cite{ferrenberg88}. If we denote the distribution of the
energy at the different coupling constants $T_i=1/(\beta_0\lambda_i)$ by
\begin{equation}
P_i(E)=P(E; T_i)
\label{eq5}
\end{equation}
we should have the identity
\begin{equation}
P_i(E;T_j)=\frac{P_i(E)\exp[(\lambda_i-\lambda_j)\beta_0E]}
{\int dE' ~P_i(E')\exp[(\lambda_i-\lambda_j)\beta_0E']}
\label{eq6}
\end{equation}
for all pairs $i$ and $j$. Note that in general this identity holds only
if the different subsystems are in equilibrium and hence it can be used
to check whether the total system is in equilibrium or not.

That for the PT run such a reweighting does indeed lead to a nice
collapse of the different $P_i(E;T_j)$ onto one master curve is shown in
Fig.~\ref{fig3}, where we plot these functions for the case $j=16$, which
corresponds to $T=3273$K, for $i=4,\ldots,15$ and $i=17,\ldots,27$. Thus
from this plot we have evidence that the system has indeed equilibrated
within the time span of the simulation.

Having checked that the algorithm does indeed allow to equilibrate
the system even at low temperatures it is of course important to see
how efficiently this is done.  One possibility to measure this is to
calculate the mean squared displacement (MSD) of a tagged particle,
\begin{equation}
\langle \Delta r^2(t) \rangle =\langle |{\bf r}_j(t)-{\bf r}_j(0)|^2\rangle \quad .
\label{eq7}
\end{equation}

\begin{figure}[bth]
\psfig{figure=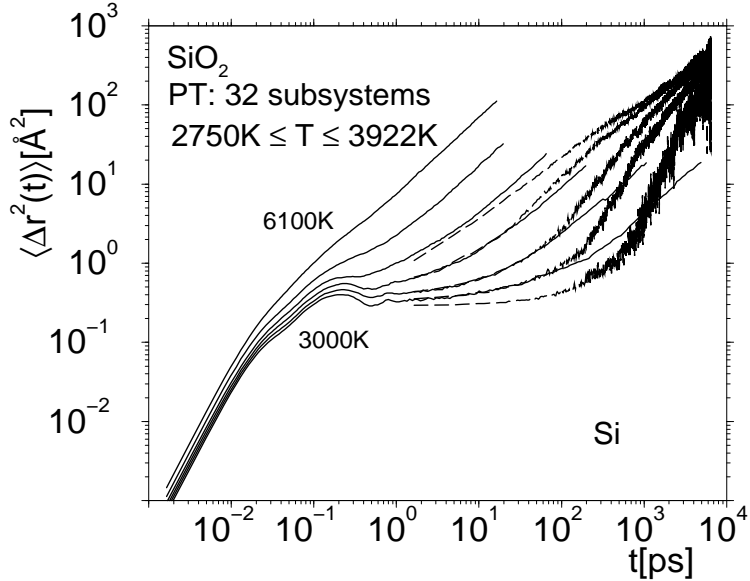,width=10cm,height=7.5cm}
\caption{Time dependence of the mean squared displacement for Si at
different temperatures. The dashed lines are from PT runs and correspond
to temperatures 3922K, 3585K, 3235K, 3019K, and 2750K (top to bottom). The
solid lines are from conventional molecular dynamics runs and correspond
to temperatures 6100K, 4700K, 4000K, 3580K, 3250K, and 3000K (top to bottom).}
\label{fig4}
\end{figure}

The time dependence of $\Delta \langle r^2(t) \rangle$ is shown in
Fig.~\ref{fig4} for the case of silicon for various temperatures (dashed
lines)~\footnote{Note that the MSD from the PT simulation was calculated
by starting at configurations that were not yet in equilibrium. Thus
the equilibrium MSD will be slightly different for times smaller than 
the equilibration
time. We have checked, however, that at long times the shown curves are
identical to the one in equilibrium.}.  Also included in the graph are
the MSD obtained from a standard microcanonical run of the same system at
similar temperatures (solid lines)~\cite{scheidler99}. From this figure
we recognize that at low temperatures and long times the MSD from the
PT is larger by about a factor to 100, thus demonstrating that this type
of dynamics is significantly faster than the conventional one.

The silica system we just considered was a glass former whose structure
is given by an open tetrahedral network and whose dynamics at low
temperatures shows an Arrhenius dependence~\cite{horbach99a}. The
Lennard-Jones system we consider next has very different properties in
that its structure is rather similar to a dense packing of hard spheres
and its dynamics shows a temperature dependence which is stronger than
a Arrhenius law~\cite{kob_lj,gleim98}.  Therefore it is of interest to
see whether the PT method also works for this kind of system, which is
done now.

\begin{figure}[bth]
\psfig{figure=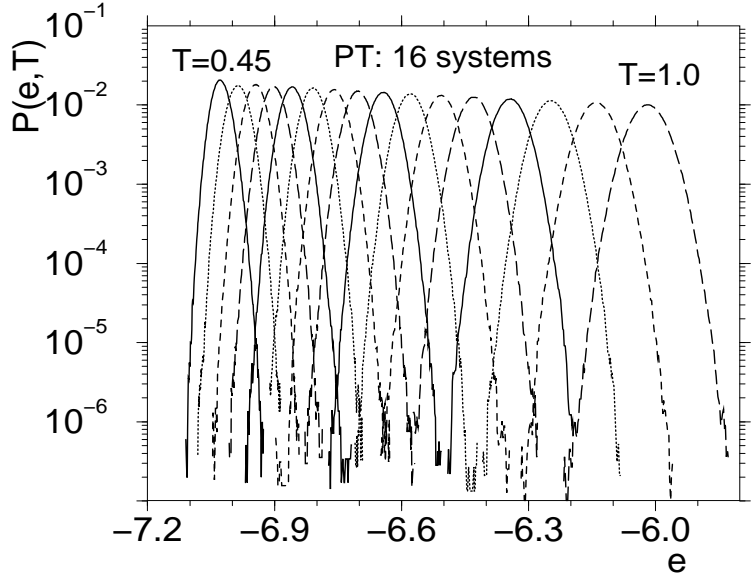,width=10cm,height=7.5cm}
\caption{Distribution of the energy for the different subsystems for the
Lennard-Jones system.}
\label{fig5}
\end{figure}

\begin{figure}[bth]
\psfig{figure=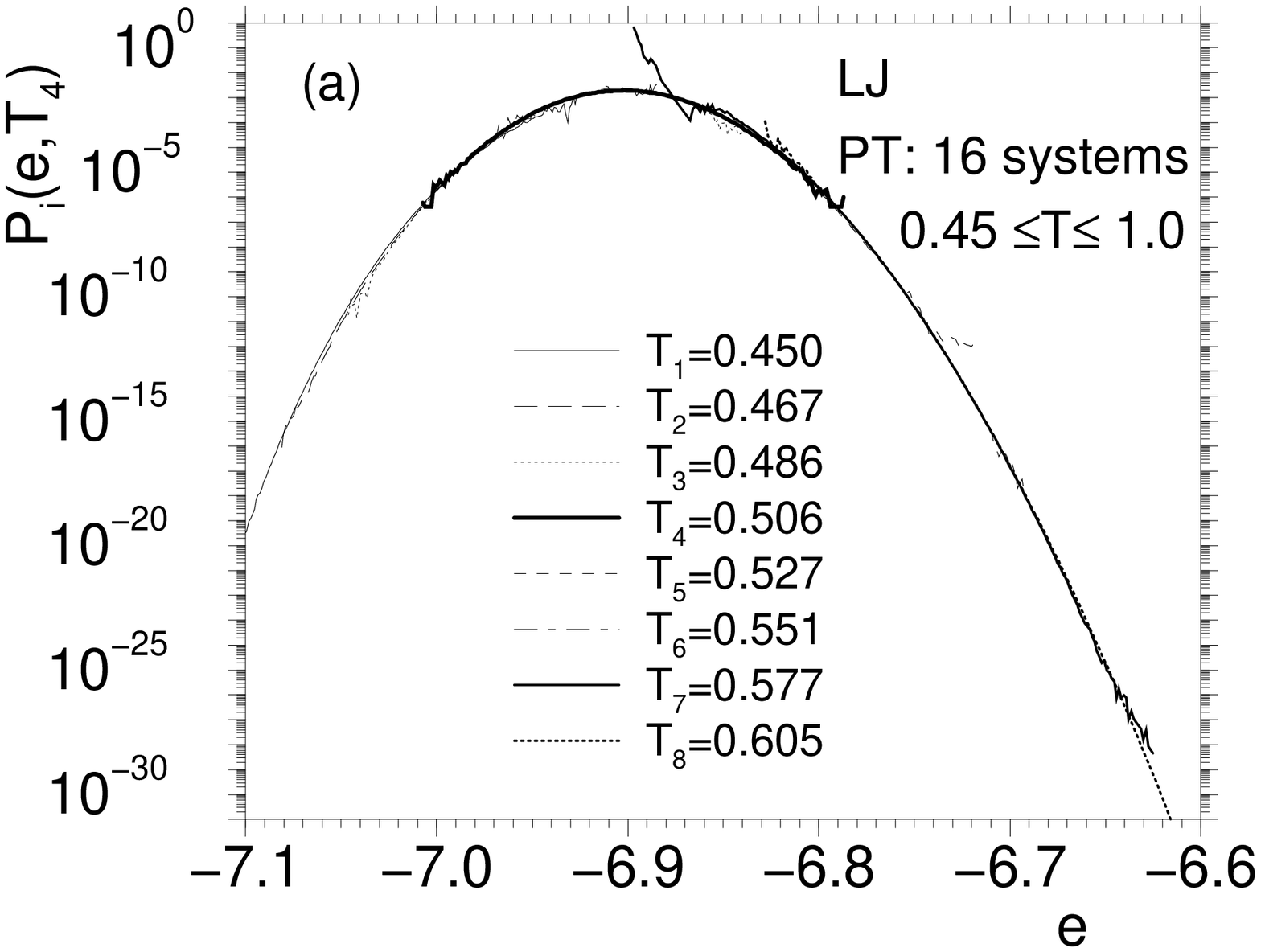,width=10cm,height=7.5cm}
\psfig{figure=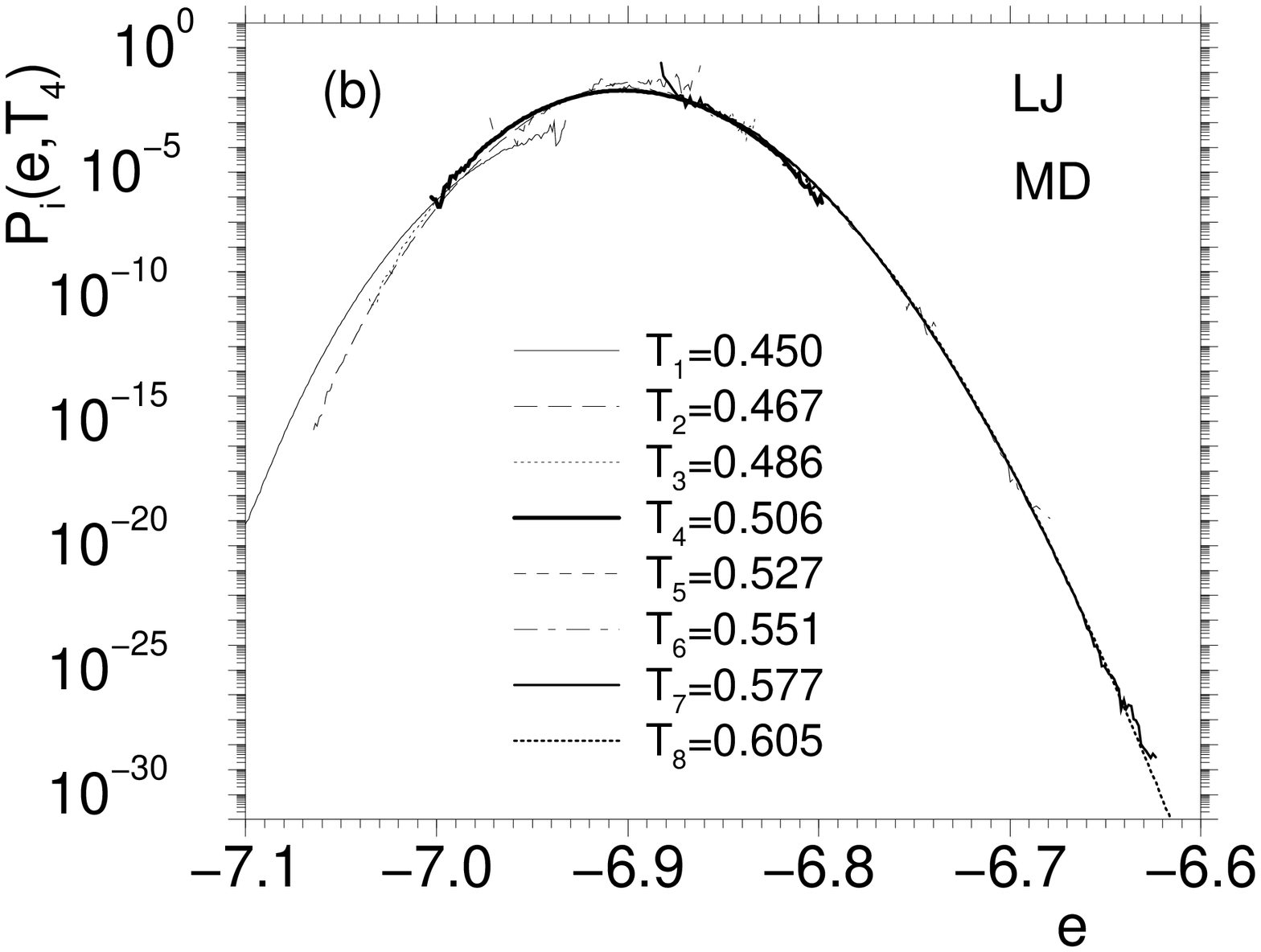,width=10cm,height=7.5cm}
\caption{Probability distribution for the energy of the Lennard-Jones
system for the temperatures given in the figure reweighted to $T=0.506$. a) data
from PT. b) data from conventional molecular dynamics.}
\label{fig6}
\end{figure}

In Fig.~\ref{fig5} we show the distribution of the energy for all 16
subsystems.  From the curves we see that in this case it is possible to
obtain much better data than for the silica system, since the numerical
demand is quite a bit smaller in the former type of system (due to
the short range nature of the interactions). That also in this case
the PT dynamics is indeed able to equilibrate the system is demonstrated
in Fig.~\ref{fig6}a where we show the same distribution functions
reweighted to the temperature $T=0.506$. From this figure we see that
all neighboring distributions collapse nicely onto the one for $T=0.506$,
thus giving evidence that the system is indeed in equilibrium. In order
to check whether this type of test is indeed sufficiently sensitive to
detect whether or not the system is in equilibrium we have also made a
run with standard molecular dynamics (at constant ``temperature" 0.45)
using as starting configuration the same as we used for the PT at that
temperature. The length of this simulation was the same as the one for the
PT and from earlier simulations it is known~\cite{kob_lj} that this time
is not sufficient to equilibrate the system via conventional molecular
dynamics. From Fig.~\ref{fig6}b one sees that the curves stemming from
the low temperatures do not fall on the master curve obtained from the
higher temperatures.  Thus we conclude that this way to analyze the data
is indeed able to detect whether or not the system is in equilibrium
and hence we have good evidence that Fig.~\ref{fig6}a shows that the PT
method has equilibrated the system.

\begin{figure}[bth]
\psfig{figure=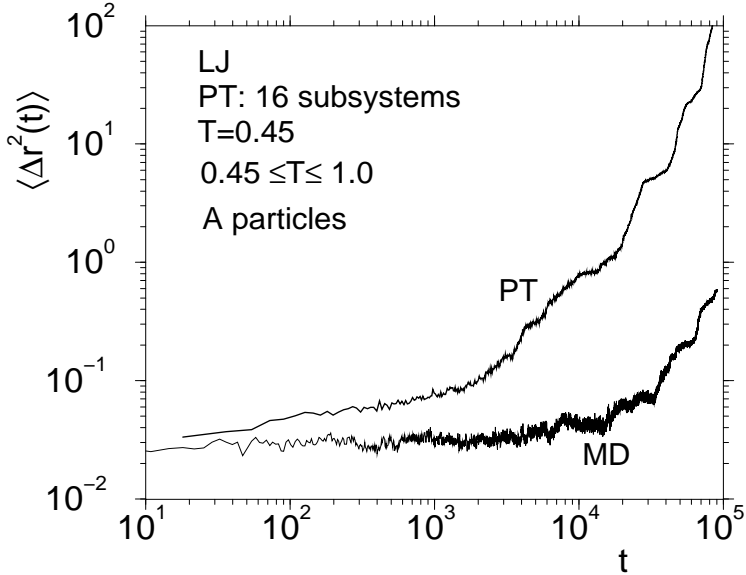,width=10cm,height=7.5cm}
\caption{Mean squared displacement of the Lennard-Jones particles at the lowest 
temperature of the PT run and the MSD from a normal molecular dynamics run.}
\label{fig7}
\end{figure}

Also in this case we judge the efficiency of the PT method by calculating
the time dependence of the mean squared displacement of the A particles
and compare it with the one obtained from standard molecular dynamics
simulations. In Fig.~\ref{fig7} we show the two curves for the lowest
temperature, $T=0.45$. It is recognized immediately that also for the
Lennard-Jones system the PT algorithm leads to a much faster propagation
of the system through configuration space in that at long times the MSD
from the PT is around 100 times larger than the one from the standard
molecular dynamics.

Finally we investigate the efficiency of the PT algorithm for the third
type of system, the fully connected Potts-glass. We have done two PT
simulations with $T=0.9$ and $T=0.7$ as the lowest temperatures. Also
for this system we have checked that the distribution functions for
the energy, $P_i(E)$,  obtained in the different subsystems can be
reweighted onto one master curve, which is again evidence that the PT
algorithm does indeed equilibrate the system. In order to see how fast this is done
we have calculated the normalized spin-autocorrelation function $C(t)$:
\begin{equation}
C(t)=\frac{1}{N(1-1/q)} \sum_i^N (\delta_{\sigma_i(t)\sigma_i(0)}-1/q)\quad.
\label{eq8}
\end{equation}

\begin{figure}[bth]
\psfig{figure=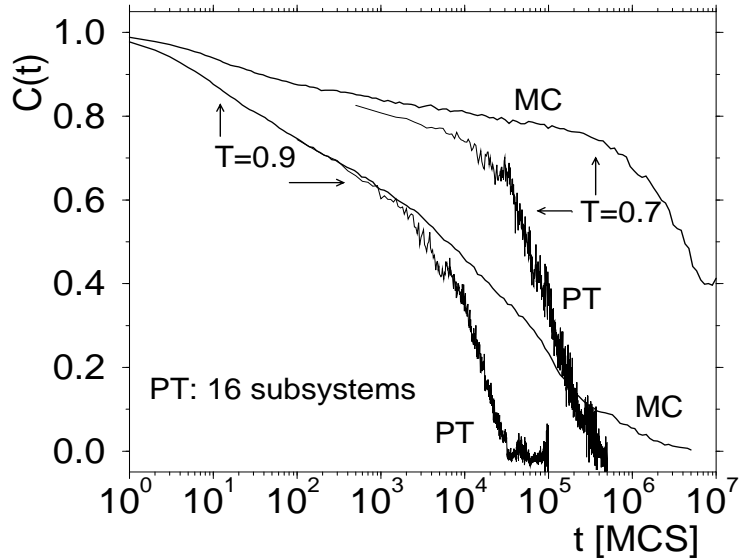,width=10cm,height=7.2cm}
\caption{Spin-autocorrelation function for the fully connected Potts-glass ($q=10$)
as obtained from the PT algorithm and the standard single spin flip Monte Carlo
Method.}
\label{fig8}
\end{figure}

The time dependence of this quantity is shown in Fig.~\ref{fig8} at
the two mentioned temperatures. Using configurations which have been
equilibrated with the PT algorithm as starting configurations, we also
did conventional single spin flip Monte Carlo simulations at the same
temperature. The resulting autocorrelation functions are included in the
figure also. From the different curves we recognize that at the higher
temperature the PT method leads to a relaxation which is more than one
decade faster than the one of the standard Monte Carlo procedure. This
factor has increased to more than 100 at the lower temperature and
we see that at this temperature the equilibration of the system with
the standard Monte Carlo method becomes hardly feasible. We have also
checked that this type of speedup is typical in that it does not depend
on the realization of the disorder, i.e. the bonds $J_{ij}$ in the
Hamiltonian~(\ref{eq4})~\cite{brangian00}. Thus we conclude that also for
this glass model the PT algorithm leads to a much quicker equilibration
of the system than a standard simulation method.

\section{Summary}
In this article we have tested to what extent the parallel tempering
algorithm is useful to equilibrate glassy systems. Whereas in the past
it has been shown that this method works well for lattice systems,
such as Ising spin systems~\cite{Hukushima,wang}, we now focused on
off-lattice models. In particular we investigated a model for SiO$_2$,
a network glass-former whose transport properties at low temperatures
show an Arrhenius dependence, and a binary Lennard-Jones system, a glass
former which is structurally similar to a hard sphere system and whose
temperature dependence of the relaxation times shows at low temperatures
a strong deviation from the Arrhenius law. We have found that for both types
of glass-formers the PT method leads to a significant ($O(10^2)$)
acceleration of the relaxation dynamics and that this factor increases
even more with increasing temperature. These results are confirmed by
our simulation of a fully connected Potts-glass, a frustrated system
which is believed to share many similarities with structural glasses.

It must be expected that there exists a temperature $T_{g,PT}$
below which also the PT method is not able to equilibrate the system,
thus hindering one to investigate the equilibrium dynamics below this
effective glass transition temperature. What the value of $T_{g,PT}$ is,
how it depends on the type of system, the system size, or the PT exchange
time $t_{PT}$, is presently not clear and thus has to be investigated
in more detail. However, already now we see that this ``new'' algorithm
(its main features dates back to 1986!~\cite{Swendsen}) will allow that
computer simulations probe the static and dynamics properties of glassy
systems at temperatures which are well below the ones that have been
accessible so far.

We thank R. H. Swendsen for helpful discussions and gratefully acknowledge
the financial support by DFG under SFB 262, the Grants in Aid for
Scientific Research from the Ministry of Education, Science, Sports
and Culture of Japan, the NIC in J\"ulich, and the Human Genome Center,
Institute of Medical Science, University of Tokyo.

\end{document}